# Imaging at depth in tissue with a single-pixel camera

Vicente Durán[1*], Fernando Soldevila[2], Esther Irles[2], Pere Clemente[2], Enrique Tajahuerce[2], Pedro Andrés[3] and Jesús Lancis[2]

[1]Microtechnology and Nanoscience Department (MC2), Chalmers University of Technology, SE 41296, Gothenburg, Sweden

[2]GROC•UJI, Institut of New Imaging Technologies, Universitat Jaume I, 12070 Castelló, Spain

[3]Departamento de Óptica, Universitat de València, 46100 Burjassot, Spain

[*]vandres@chalmers.se

One challenge that has long held the attention of scientists is that of clearly seeing objects hidden by turbid media, as smoke, fog or biological tissue, which has major implications in fields such as remote sensing or early diagnosis of diseases. Here, we combine structured incoherent illumination and bucket detection for imaging an object completely embedded in a turbid medium. A sequence of low-intensity microstructured light patterns is launched onto the object, whose image is accurately reconstructed through the light fluctuations measured by a single-pixel detector. Our technique is noninvasive, does not require coherent sources, raster scanning nor time-gated detection and benefits from the compressive sensing strategy. We experimentally retrieve the image at visible wavelengths of a transilluminated target embedded in a 6mm-thick sample of chicken breast.

# 1. INTRODUCTION

Biological tissues present a high degree of transparency for near infrared (NIR) light, since natural chromophores do not significantly absorb radiation within such a spectral range. At irradiance doses below the tissue damage threshold, the penetration depth of NIR light can be of several centimeters. Certain optical diagnostic modalities take advantage of the above 'diagnostic window' for the quantitative evaluation of pathological structures[1]. Nevertheless, tissues are characterized by a strong scattering of NIR radiation, which prevents one from achieving deep-tissue imaging. Photons propagating through an inhomogeneous medium can experience a certain number of scattering events and, as a result, the emerging radiation is a mix of two components: "image-bearing" photons, travelling nearly in the incoming direction, and diffusive photons, which have experienced multiple scattering inside the medium and follow a completely uncorrelated direction at the output.

Modelling the propagation of diffusive photons enables some hybrid imaging techniques to penetrate deeply into a tissue ($> 1\ cm$)[2,3]. However, the low resolution offered by these methods restricts them to imaging complete organs in living beings. The most widespread principle for high-resolution imaging through turbid media is the isolation of the unscattered photons. This idea is at the heart of so diverse approaches such as scanning multiphoton microscopy[4] or imaging techniques based on time-resolved[5], coherence-gated[6] and polarization-sensitive detection[7]. The intensity of the unscattered light component falls off with the medium thickness. As a consequence, imaging begins to be difficult or unfeasible at a penetration depth larger than the tissue transport mean free path (TMFP), $l_t$, a measurement of the mean distance that photons travel before they completely 'forget' their initial direction[1]. Typical values of this parameter are of the order of 1 mm. This barrier can be beaten by tissue optical clearing techniques, but they entail the immersion of tissues into chemical agents[8,9].

Recently, it has been demonstrated a non-invasive approach that uses diffusive photons while preserving optical resolution. This method is based on the angular correlation ("memory effect") inherent to the speckle patterns generated by photon scattering[10]. However, present implementations are restricted to submillimeter-thick tissue[11], since the memory-effect range is inversely proportional to the medium thickness. In this paper, we present a novel "scattering-free" imaging approach, based on the concept of the single-pixel camera[12,13,14,15,16] which enables to image objects embedded at a depth of several times the tissue TMFP. Our incoherent low-intensity technique uses forward-scattered photons at the illumination stage, but takes advantage of the total flux of the photons at the detection step. We capture the object information with a "bucket" detector without spatial resolution, responsible for measuring light fluctuations that are correlated with a set of light patterns sequentially projected onto the object.

## 2. SINGLE-PIXEL IMAGING INSIDE SCATTERING MEDIA

Our single-pixel scheme is sketched in Fig. 1. Spatially and temporally incoherent light coming from a white-light source impinges onto a digital micromirror device (DMD), which is composed of an array of electronically controlled micromirrors that can tilt between two angular directions. The DMD produces a set of illumination patterns, which are projected by an optical relay system onto a high-contrast object embedded in a non-absorbing inhomogeneous medium. At the mesoscopic scattering regime[17], i.e., for depths inside the interval 1-10 $l_t$, the light that hits the object consists of two superimposed components: a strong diffusive halo, the result of averaging many uncorrelated and noninterfering speckle patterns generated by the incoherent source, plus a forward-scattered weak signal, which is a "ghost" illumination pattern with a spatial structure similar to the light pattern "sculpted" by the DMD. Finally, in our transillumination geometry, the light emerging from the back portion of the turbid medium is concentrated by a collecting lens onto the large active surface of a bucket detector, here a photodiode (PD).

The purpose of the above optical system is to obtain the correlation between the weak structured patterns sampling the object and the tiny fluctuations that appear in the bucket signal. Such a correlation enables to get rid of the prevalent noisy background interacting with the object due to scattering. As the DMD projects onto the object a sequence of $M$ masks $\{I_i(\boldsymbol{x})\}$ ($i=1,…M$), the intensity at each position $\boldsymbol{x}$ of the object fluctuates from one pattern to the next. On its turn, the bucket detector measures a signal $\{Y_i\}$, where each $Y_i$ is a photocurrent proportional to the total light power emerging from the sample for the $i$-th mask. The image $T(\boldsymbol{x})$ is then built from the spatially resolved cross-correlation $T(\boldsymbol{x}) = \sum_i \Delta I_i(\boldsymbol{x})\, \Delta Y_i$ between the pattern-to-pattern intensity fluctuations, $\Delta I_i = I_i(\boldsymbol{x}) - \langle I(\boldsymbol{x}) \rangle$, and those registered by the bucket detector, $Y_i - \langle Y \rangle$. Interestingly, for a given illumination pattern, the bucket detector provides a weighted-sum of the fluctuations arising from all the pixels of the pattern and the weights of this sum correspond to the object´s transmission at each spatial position. Hence, when the pattern-to-pattern fluctuations that occur at each pixel are correlated with the photocurrent signal provided by the bucket detector, the strength of the correlation is proportional to the object's transmission at the corresponding position. This result is not affected by the diffusive background, since it is uncorrelated with the "ghost" structured pattern that impinges onto the object.

Our approach can be reformulated into the framework of computational single-pixel imaging, where the $N$-pixel image $T$ of an object is recovered from its $M$ incoherent projections in a proper function basis. This formulation leads to the algebraic problem $Y = \Theta T$, where $\Theta$ is a matrix constructed by arranging in successive rows each illuminating pattern. To form a completely determined set of measurements, the rank of $\Theta$ must equal the object's data dimension. Our procedure requires the sequential delivery of *N* illumination patterns, so the refreshing rate of the DMD limits the system temporal resolution (i.e., the time necessary to acquire one image). To speed up the data acquisition, we can take advantage of the ground-breaking theory of compressive sensing (CS), which makes it possible to recover an *N*-pixel image from *M<N* measurements.[18] Another aspect to be considered is the choice of the multiplexing masks, key to achieve a high-fidelity imaging. We implement patterns derived from the Hadamard matrices[16], which have been proven to improve by a factor $\sqrt{N}$ the signal-to-noise ratio (SNR) provided by standard raster-scanning[19]. In addition, the efficiency of our scheme benefits from the use of a single-element detector to simultaneously detect all the photons transmitted by the object, instead of spreading them out over an array of sensors in a pixelated device.

Detection is of paramount importance for deep-tissue imaging. In time-gating or coherence imaging techniques, the useful signal is composed exclusively of "image-bearing" photons arriving to the detector, which drastically limits the maximum total thickness of the tissue[17]. In our approach, "image-bearing" photons that are transmitted by the object become diffusive as they migrate through the back inhomogeneous layer (Fig. 1b). The single-pixel detection of diffusive photons has demonstrated to be crucial for image transmission through scattering media, even when they are dynamic[20]. Here we go one step further, since the contribution of diffusive photons to the signal fluctuations leads to non-invasive imaging through tissue, although at the expense of enlarging the detection field of view, a situation that resembles that found in two-photon microscopy[4].

## 3. METHODS AND EXPERIMENTAL RESULTS

We used the experimental setup shown schematically in Fig. 1a. The light source was a high-power Xe lamp (Model 66002, Oriel Corporation). The SLM was a DMD with a pixel pitch of 10.8 $\mu$m (DLP Discovery 4100, Texas Instruments). The unit cell of the binary intensity patterns displayed onto the modulator was composed of 8×8 DMD pixels. Two thin lenses projected the illuminating patterns onto the object and enlarged their size with a lateral magnification of 1.9, so the length of the unit cell became 167 $\mu$m. An optical collecting system formed by a high NA photography objective (AF Nikkor 50 mm f/1.4D, Nikon) and a condenser lens (25.4 mm × 25.4 mm FL Condenser Lens, Edmund) ensured that the light was focused within the sensor area (13 mm²) of a photodiode (DET36A Si Biased Detector, Thorlabs). An analog-to-digital converter (PRO8000 with a PDA8000 module, Thorlabs) digitized the photodetected signal. Custom software written in LabVIEW was used to control the measurement process. Temporal averaging of the photodetected signal was used to improve its SNR, which limited the projection rate of the Hadamard patterns to 5 Hz.

To assess the capability of our technique for tissue imaging, we retrieved an image of a binary amplitude object (Fig. 2a) sandwiched between two layers of chicken breast (Fig. 2b), with a thickness of 2.84 mm and 2.92 mm, respectively. Accepting that $l_t = 1\ mm$ for chicken breast at green wavelengths[21], the object was placed at a depth of about 3$l_t$, clearly within the mesoscopic regime. The image reconstruction is presented in Fig. 2c. A sample of the measured light fluctuations, corresponding to the first 500 Hadamard patterns, are shown in Fig. 2d. Such fluctuations make it possible to reconstruct the object image, as can be seen by comparing them with those obtained with the object immersed in air (Fig. 2d). To be detectable, the induced fluctuations must be at least higher than the shot-noise of the detector. Indeed, quantum noise, as well as the tissue damage threshold, ultimately limits the penetration depth of our technique.

The effect of decreasing the illumination level on the image quality can be observed in Fig. 3. For this experiment, the object shown in Fig. 2a was hidden by an inhomogeneous layer of a fixed thickness. For intensities of a few tens of µW/cm$^2$, the peak SNR (PSNR) corresponds to images of good quality (PSNR>20 dB). As the illumination level decreases, a growing number of intensity fluctuations (and, hence, of image coefficients) are hidden by the detector noise, leading to a rapid PSNR falloff. The relevance of above results for imaging at depth is highlighted when we compare the intensities involved in Fig. 3 with, for example, the skin damage threshold for a continuous wave source[22] (~0.2 W/cm$^2$). Remarkably, this value is three orders of magnitude higher than the maximum input intensity attainable with our setup (135 µW/cm$^2$).

The resolution of our single-pixel scheme was analyzed with the NBS 1963A resolution chart. Figs. 4a-b show, respectively, images of the same group of bars when the object was immersed in air and after hiding it behind a 2.4-mm-thick layer of chicken breast. To quantify the resolution decrease in the second case, we removed the tissue and searched a group of bars of higher line frequency that provided the same contrast as that measured in Fig. 4b. As the line frequency was doubled (Fig. 4c), the presence of the tissue made the image resolution to be reduced by a factor 2. Since resolutions lower than 2$\mu$m could be achieved by our system[23], objects of a few microns hidden by millimeter tissue are expected to be resolved, a promising achievement considering the trade-off existing between penetration depth and resolution[17]. However, our approach is currently limited to high-contrast objects and the use of mechanisms of optical contrast, mainly fluorescence, is yet to be tested.

## 4. CONCLUSIONS

In conclusion, we have presented a technique that merges Hadamard illumination and single-pixel photodetection for tissue imaging at a penetration depth of 3$l_t$ with visible light. In our approach, images free of scattering noise are reconstructed by detecting intensity fluctuations with a light exposure much lower than the tissue damage threshold. The above goal has been reached without increasing the cost and complexity of the imaging system, as is evidenced by the use of a white-light lamp and an off-the-shelf DMD. Image resolution and field of view are controlled by the DMD parameters and the data acquisition process can be sped up by the application of compressive sensing. In addition, the freedom to select a proper bucket detector can lead to performing hyperspectral or polarimetric imaging[24]. Our results can be considered as a technology-enabled version of the first experiments in mammography conducted by doctors in the Victorian age using candle light[25].

**Figures**

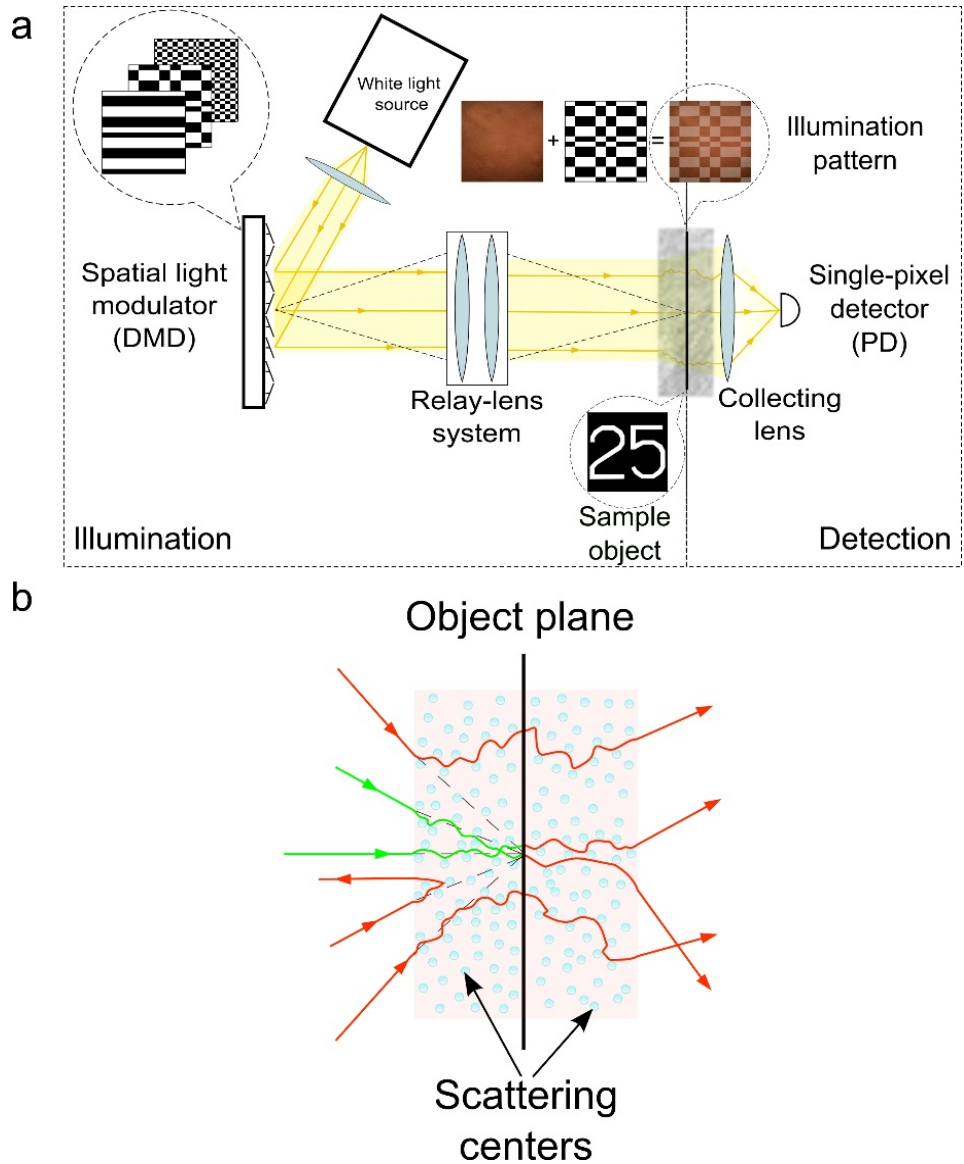

**Fig. 1.** Operation principle. (a) Schematic of the optical setup. Upper left inset: examples of projected patterns. Upper right inset: weighted superposition of the diffusive background and the illumination pattern. The contribution of the latter has been artificially increased to make it visible. Lower inset: binary amplitude object. (b) A ray light representation corresponding to a bright point of an illumination pattern. Photons are randomly deviated from the directions they would trace in a homogeneous medium (dashed lines), forming a diffuse halo on the object plane (red rays). However, a fraction

of them follows a zigzagging path close to the dashed one, even after several interactions with the scattering centres (green rays). If such rays are transmitted by the object, they eventually become red rays as a consequence of the second scattering process. Detection consists of collecting all the rays emerging from the ensemble in any direction.

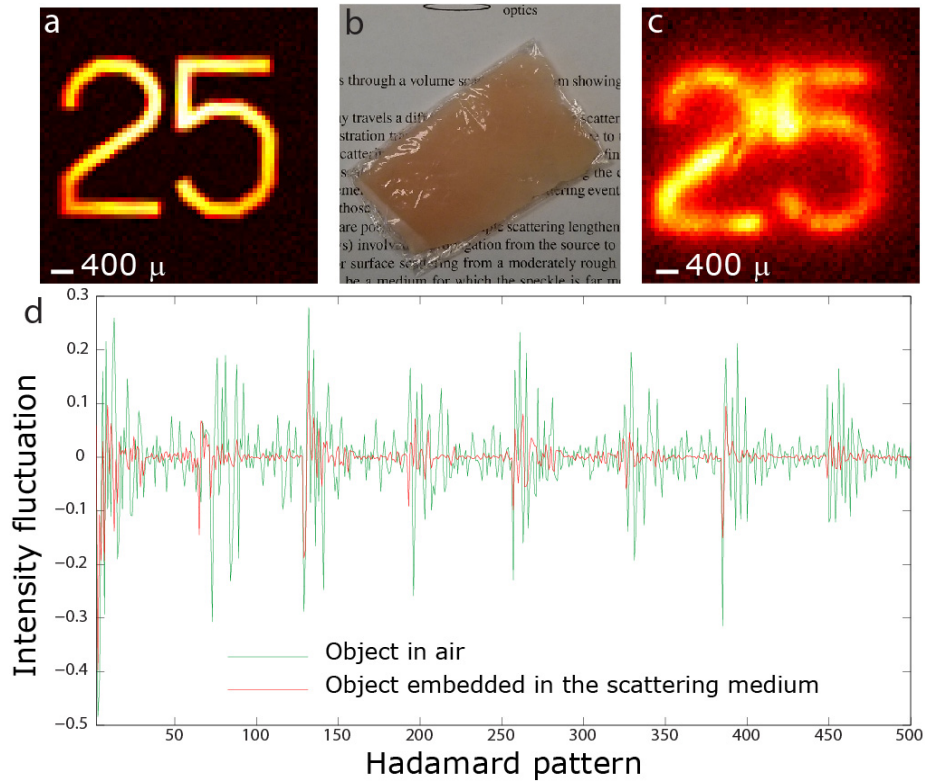

**Fig. 2.** Imaging inside a biological tissue: experimental results. (a) Single-pixel reconstruction of a binary amplitude object, a selected area of the NBS 1963A resolution chart containing the digits 25. For this reconstruction, the object was immersed in air. (b) Photograph of one of the 3-mm samples of chicken breast that were used in this experiment. (c) Image reconstruction (64×64 pixels) when the object was embedded in chicken breast. (d) Normalized intensity fluctuations corresponding to the first 500 Hadamard patterns used to recover the object in (a) and (c).

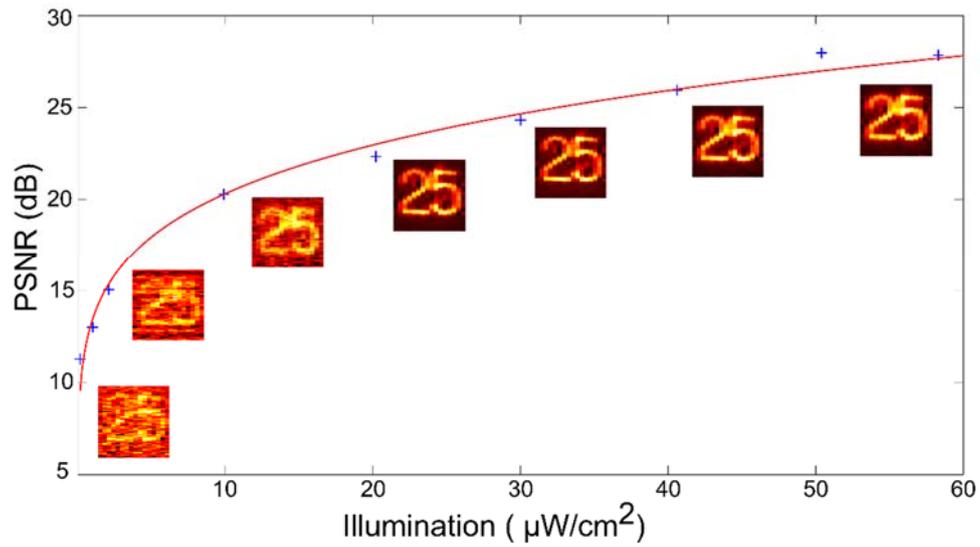

**Fig. 3.** The effect of decreasing the source intensity. PSNR of image reconstructions obtained with different illumination levels. For $I_{input}<10^{-5}$W/cm$^2$, the reduction in the image quality is evident (PSNR<20 dB). The plot includes some of the reconstructions (each one of 64×64 pixels) to illustrate the evolution of the image quality.

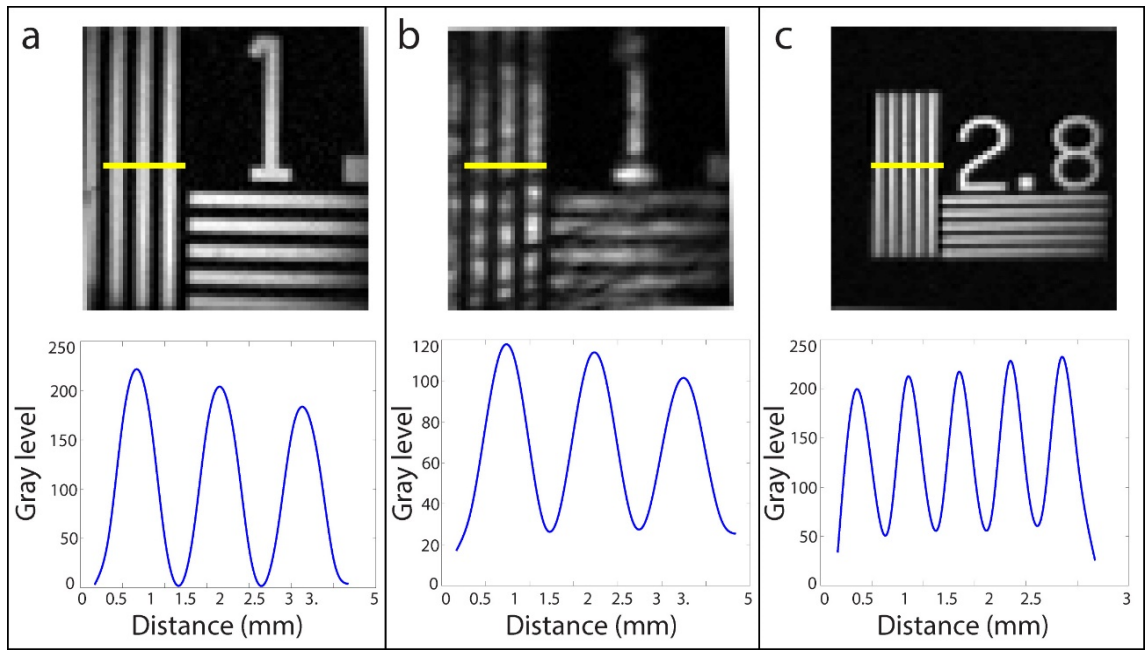

**Fig. 4.** Imaging system resolution. (a), (b) Reconstructed images of a group of bars (1.4 cycles/mm) when the resolution target is, respectively, immersed in air and hidden by a 2.4-mm-thick layer of chicken breast. Horizontal profiles to calculate the contrast of the vertical lines are also included. To reduce noise artifacts, a set of image rows around the marked yellow lines were averaged and the resulting curves were smoothed. (c) Reconstructed image of another group of bars (2.8 cycles/mm) without any tissue hiding the target. The corresponding horizontal profile has a contrast similar to the one calculated in (b).